\documentstyle[12pt]{article}
\begin{document}
 
\title{ Baryon-Baryon Interactions }
 
\author{ Th. A. Rijken \\[0.5cm]
Institute for Theoretical Physics, \\[0.3cm]
University of Nijmegen, The Netherlands }
\date{}
\maketitle
 
\abstract{
After a short survey of some topics of interest in the study of
baryon-baryon scattering,
the recent Nijmegen energy dependent partial wave analysis (PWA)
of the nucleon-nucleon data is reviewed. In this PWA
the energy range for both $pp$ and $np$ is now $ 0 < T_{lab} < 350$ MeV
and a $\chi^{2}_{d.o.f.}=1.08$ was reached.
The implications for the pion-nucleon coupling constants
are discussed. Comments are made with respect to recent discussions
around this coupling constant in the literature.
In the second part, we briefly sketch
the picture of the baryon in several, more or less QCD-based,
quark-models that have been rather prominent in the literature.
Inspired by these pictures we constructed a new soft-core model
for the nucleon-nucleon interaction and present the
first results of this model in a $\chi^{2}$-fit to the new
multi-energy Nijmegen PWA.
With this new model we succeeded in narrowing the
gap between theory and experiment at low energies. For the energies
$ T_{lab} = 25-320$ MeV we reached a record low
$\chi^{2}_{p.d.p.}=1.16$.
We finish the paper with some conclusions and an outlook
describing the extension of the new model to baryon-baryon
scattering. }
 
\section{Introduction}
A review of baryon-baryon scattering and the early work by the
Nijmegen group has been given in \cite{SNRV71}. Reviews of the recent
work can be found in {\it e.g.} \cite{Padua89} and \cite{LA90}.
For the nucleon-nucleon work of other groups, like Bonn and Paris,
we refer the reader to \cite{Machleidt87}.
 
\noindent
Although the items we discuss here are relevant, directly or indirectly,
for all baryon-baryon channels, we focus in this paper mainly on the
nucleon-nucleon channels. A shopping list of the items about
which we want to learn more through the analysis of the
experimental data and the study of theoretical models contains for
example the following subjects:
\begin{enumerate}
\item  Long-, intermediate-, and short-range mechanisms: {\it e.g.}
        single meson-echange $( \pi, \rho, ...)$,
        double meson-exchange $(\pi\otimes\pi,\pi\otimes\rho, . . . )$,
        quark effects.
\item   Relativistic effects: {\it e.g.}
        off-energy-shell effects, off-mass-shell effects.
\item   Chiral-symmetry and soft-pion effects.
\item   $SU(2,I)$- and $SU(3,F)$-symmetry of the coupling constants.
\end{enumerate}
In this paper we concentrate on the first subject {\it i.e.}
the mechanisms behind the nuclear force. Now, it is well known that
the theoretical models do not explain the $N\!N$-data better than
with a $\chi^{2}_{p.d.p.} \geq 1.8$. In this paper we describe
a first attempt to investigate whether the new Nijmegen multi-energy
partial wave analysis (PWA) \cite{SKRS93} allows a better theoretical
description of the data. This is done by an extension of the Nijmegen
soft-core model \cite{NRS78}.
 
The contents of this paper is as follows. In section 2 we report on the
Nijmegen $pp + np$ multi-energy PWA. In section 3 we review briefly the
situation around the pion-nucleon coupling constant. In section 4 we
list the popular quark models and emphasize the synthesis of these
quark models and the non-relativistic quark model in the general
physical picture of a baryon as advocated by the chiral-quark model.
In section 5 we introduce together with its first results,
a new soft-core model, which we henceforth call the
extended-soft-core (ESC) model.
Finally, in  section
6 we offer some conclusions and an outlook. Here we indicate how the
ESC-model can be extended to all baryon-baryon channels.
 
\section{Multi-Energy Partial-Wave-Analysis}
After the multi-energy phase shift analysis of the $pp$ data below
350 MeV \cite{Berg90}, the Nijmegen group has recently finished a
similar analysis for the $pp$ and $np$ data \cite{SKRS93}.
The $pp + np$ data base consists of 1787 $pp$-data and 2514 $np$-data.
The principal method employed in this multi-energy PWA
consists in a division of the internucleon distances
$r_{N\!N}$ into three regions:
\begin{enumerate}
\item[(i)] $r_{N\!N} \geq 2.0 $ fm: the long-range region. Here the
potential $V=V_{L}$ is dominated by the well known electromagnetic
and one-pion-exchange potentials, $V_{L} \approx V_{EM} + V_{OPE}$.
The residual potential comes from the spurs of the HBE, see next item.
\item[(ii)] $b \leq r_{N\!N} \leq 2.0 $ fm ($b = 1.4$ fm):
the intermediate-range region. Here the potential is taken to be
a sum of the one-pion-exchange (OPE)
and the heavy-boson exchanges (HBE) from
the Nijmegen \cite{NRS78} soft-core potential, so
$V= V_{EM} + V_{OPE} + V_{HBE}^{N}$.
For the singlet waves the following modification
proved to be advantageous:
$V_{HBE}^{N} \rightarrow f^{s}_{med} V_{HBE}^{N}$, with
$f^{s}_{med}=1.8$.
\item[(iii)] $ r_{N\!N} \leq b $ fm:
the short-range region. Here an energy dependent boundary condition
is used in principle. In practice it appeared useful to use energy
dependent square well potentials in the inner region. This is equivalent
to
 \[ P\left(b;k^{2}\right) =
  P_{free}\left(b;k^{2}-2M_{r} V_{S}\right)   \]
 The parametrization of the energy dependence is as follows
 \[
    V_{S,\beta}(k^{2}) = \frac{1}{2M_{r}} \sum_{n=0}^{N}
      a_{n,\beta}\ k^{2n}  \]
 independently for each wave $\beta=(L,S,J)$. Here, $k$ denotes the
 relativistic cm momentum. For each wave only a
couple of 'phase parameters' $a_{n}$'s were needed to cover the energy
interval $ 0 \leq T_{lab} \leq 350$ MeV unbiased. In total
21 phase parameters were used for $pp$ and 18 for $np$.
\end{enumerate}
With the parametrization of the potentials completed, the
radial Schr\"{o}dinger equation
 \[
  \left(\frac{d^{2}}{dr^{2}} + k^{2} - \frac{L^{2}}{r^{2}} -
     M_{r} V_{\beta}(r) \right) \chi_{\beta}(r) = 0  \]
is solved and the phase shifts as a function of the parameters and the
energy are obtained.
 
Very important ingredients of this PWA are:
\begin{enumerate}
\item[a.] The accurate treatment of the electromagnetic interactions:
\[
  V_{EM} = \tilde{V}_{C}\ +\ V_{MM}\ +\ V_{VP}  \]
where $\tilde{V}_{C}$ is the improved Coulomb interaction,
$V_{MM}$ is the magnetic moment interaction, and
$V_{VP}$ is the vacuum polarization.
\item[b.] The OPE-amplitude is treated in
Coulomb-distorted-wave Born-approximation.
 It appeared that a simple Coulomb barrier penetration
 factor was not sufficiently realistic. This CDWBA-treatment is very
important in the determination of the pion-nucleon coupling constant.
\item[c.] The correction of the $I=1$ $np$-waves for the
$\pi^{\pm}-\pi^{0}$-mass difference.
\end{enumerate}
 As a result of this PWA a $\chi^{2}_{d.o.f} = 1.08$ was reached.
For $pp$: $\chi^{2}=1787.0$, $N_{d.o.f}=1613$ and for
$np$: $\chi^{2}=2484.2$, $N_{d.o.f}=2332$.
In a combined $pp+np$ analysis
one obtained  $\chi^{2}=4263.8$, $N_{d.o.f}=4301$. As an indication of
the realistic energy dependence, it was found that extrapolation to
the deuteron pole results in a predicted binding energy
$B= 2.2247(35)$, whereas experimentally $B = 2.224575(9)$.
As a result of this PWA very accurate $I=0$ $np$ phases are now
available, the estimated errors are only slightly larger than those
for the $pp$ phases. The mixing parameter $\epsilon_{1}$ is not small
and reaches $4.57 \pm 0.25$ degrees at $T_{lab}=350$ MeV.
 
The Nijmegen group has also constructed Reidlike phenomenological
potential models \cite{SKTS94},
which fit the data equally well as the PWA,
{\it i.e.} $\chi^{2}_{p.d.p} \approx 1.0$. These potentials make
the results of this new PWA available for many applications in few
body systems. As an example, we mention the very recent calculation of
the triton using these Reidlike potentials \cite{FPSS93}. It was found
that these two-nucleon interactions predict the binding energy as
$7.62-7.72$ MeV.
 
\section{Pion-Nucleon Coupling Constant}
The first accurate determination of the neutral pion-nucleon coupling
constant was done by the Nijmegen group \cite{Berg87,Berg90}.
When this author presented the Nijmegen determination of this
pion-nucleon coupling constant at the Vancouver conference in 1989
\cite{RSKKS90} it was suggested that also the charged pion-nucleon
coupling constant should be determined with the same method. With the
Nijmegen 1993 PWA \cite{SKRS93}
this has been done and also the charged pion-nucleon
coupling turns out to be significantly lower than that found in the
Karlsruhe 1980-analysis \cite{Karlsruhe80}.
Meanwhile, also in a recent pion-nucleon partial wave analysis
by the VPI\&SU group a value consistent with the Nijmegen
determination was found \cite{VPI90}.
Moreover, a PWA of the combined  $pp$ and $np$ data \cite{KSS91} and
of the antinucleon-nucleon data revealed the same result \cite{Timm91}.
In \cite{STS93} the recent determinations of the $\pi N\!N$ couplings
are tabulated. Below in Table~\ref{table1} we show Table I of
 ref.~\cite{STS93}.
Here, DR refers to the use of dispersion relations, PWA to the usual
phase shift or partial wave analysis.
\begin{table}[hbt]
\begin{center}
\vspace{0.5cm}
\begin{tabular}{ccccc} \hline \hline  & & & & \\
  Group  &  Year & Method & $10^{3} f_{pp\pi^{0}}^{2}$ &
 $10^{3} f_{c}^{2}$ \\ & & & & \\
\hline    & & & & \\
 Karlsruhe-Helsinki \cite{Karlsruhe80}
 & pre-1983 & $\pi^{\pm}$ DR & & 79(1) \\
    & &  & & \\
 Nijmegen \cite{Berg90} & 1987-1990 & $pp$ PWA & 74.9(0.7) & \\
    & &  & & \\
 VPI\&SU \cite{VPI90} & 1990 & $\pi^{\pm}p$ DR &  & 73.5(1.5) \\
    & & & &  \\
 Nijmegen \cite{KSS91} &
 1991 & combined $N\!N$ PWA & 75.1(0.6) & 74.1(0.5) \\
    & & & &  \\
 Nijmegen \cite{Timm91} & 1991 & $\bar{p}p$ PWA &  & 75.1(1.7) \\
    & & & &  \\
 Nijmegen \cite{STS93} &
 1992 & $pp$ and $np$ PWA & 74.5(0.6) & 74.8(0.3) \\
 & & & & \\ \hline \hline
\end{tabular}
 \caption{ Recent $\pi N\!N$-coupling constant determinations.}
\label{table1} \end{center} \end{table}
Soon after the publication of the Nijmegen $\pi^{0}$-coupling
constant determination
\cite{Berg87}, it was heavily criticized in the literature. Notably,
the claim was made, see for example \cite{TH89},
that the Nijmegen group had overlooked form factor
effects. Still recently, it was suggested in the panel discussion
of the Adelaide conference \cite{Ericson93} that the value of the
pion coupling constant found in the Nijmegen method depends on the
shape of the form factor. This was dismissed in a Nijmegen paper
on the several issues raised in the literature \cite{STS93}. The
main points made here are:
\begin{enumerate}
\item[(i)] The Nijmegen PWA is statistically impeccable. The criteria
used in selecting the data base are unbiased and common practice under
specialists on the $N\!N$ phase shift analysis.
\item[(ii)] Tests show that indeed the Nijmegen method determines the
pole value of the pion-nucleon coupling constant.
\item[(iii)] Neither the shape nor reasonable values of the cut-off
mass have any influence.
\item[(iv)] The presently available potential models are too bad to
determine $f_{\pi N\!N}$ with an accuracy comparable to the
Nijmegen $N\!N$ phase shift analysis.
\end{enumerate}
 
\section{Baryon Structure, Chiral Quark-Models}
\begin{figure}[htb]
\vspace{12cm}
\includegraphics{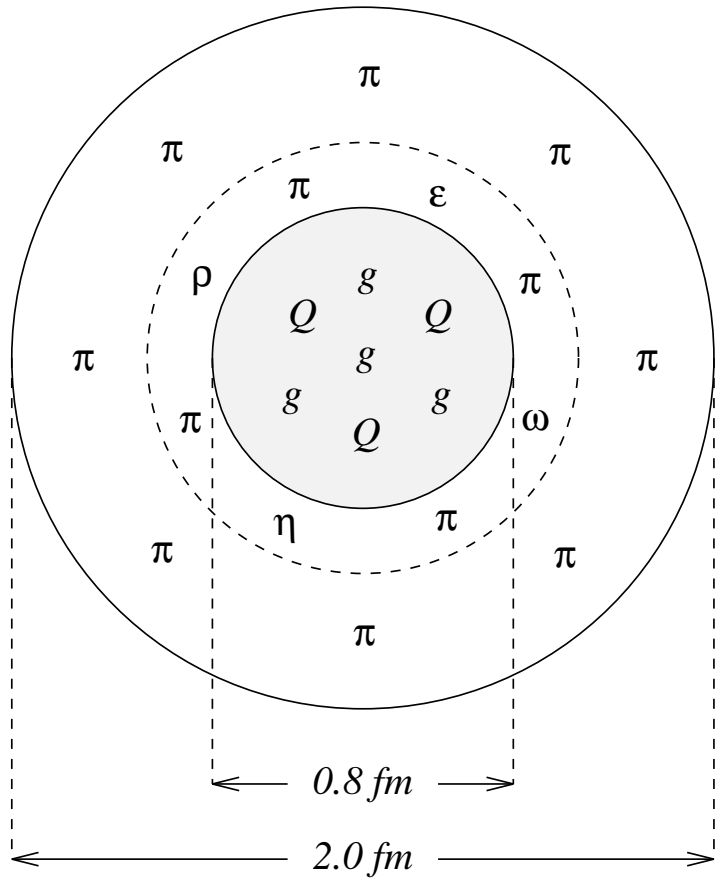}
\caption{Schematic model of the baryon structure }
\label{fig1}
\end{figure}

The quark-model picture of the baryons should be of some
directional value in the
deduction of a realistic model for baryon-baryon scattering. Interesting
bag-models are the MIT \cite{Chodos74}, the Stony Brook \cite{Brown79},
and the TRIUMF \cite{Theberge80} models.
A particularly interesting quark-model is the chiral-quark-model
\cite{Manohar84}. This model explains the successes of the
non-relativistic quark-model (NRQM) and at the same time is closely
connected with the description of hadron dynamics through
interactions involving mesons and baryons using effective chiral
lagrangians. The general idea is that the QCD-vacuum becomes unstable
at $Q^{2} \leq \Lambda^{2}_{\chi SB} \approx (1 GeV)^{2}$. The vacuum
goes through a phase transition, making for the quarks
$ \langle 0| \bar{\psi} \psi |0\rangle \neq 0$ and
the gluon coupling $\alpha_{S}$ small.
This generates the constituent quark masses and implies that
the quarks move around in the core of a baryon essentially as being
free, just as in the NRQM. Viewing (part of) the pion as the Goldstone
boson, correlated with spontaneously broken chiral invariance, makes it
natural that there is a soft-pion cloud around a constituent quark.
High energy experiments indicate that the Pomeron
couples to the quarks \cite{Henkes92}. Then,
a soft pion cloud around a constituent quark
offers a natural explanation for the multi-peripheral component of
the Pomeron. Also, the coupling of mesons to quarks dressed by a pion
cloud is in accordance with the ideas that the non-linear sigma-model
is relevant for the description of hadronic interactions \cite{Witten79}.
We have drawn for a nucleon in Fig.~\ref{fig1}
the picture that emerges from the bag-models and the
chiral-quark-model, {\it i.e.} a quark-core surrounded by a meson
cloud of pions and other mesons,
Baryon-baryon scattering is the
quantum mechanical scattering of two of such systems.
The chiral-quark-model in particular, provides a natural basis
for an approach to baryon-baryon scattering using only mesonic degrees
of freedom in the derivation of the baryon-baryon interactions. In the
next paragraph we will describe such an attempt. We construct a new
$N\!N$-model and make a fit to the 1993 multi-energy Nijmegen PWA.

\section{Extended Soft-Core model}
The potential of this new $N\!N$-model, henceforth referred to as
the ESC-model, consists of the contributions of
\begin{enumerate}
\item[(i)] The OBE-potentials of \cite{NRS78}, which apart from the
low lying pseudo-scalar-, vector-, and scalar-mesons includes also
contributions of the Pomeron. The latter represents the
multi-peripheral (soft)pion exchanges and multi-gluon exchanges.
\item[(ii)] The $2\pi$-potentials as given in \cite{Rijken91}. These are
two-pion-exchange potentials based on the pseudo-vector pion-nucleon
coupling. We include only the so-called BW-graphs, {\it i.e.} we
discard the TMO-graphs (see \cite{Rijken91} for this nomenclature).
This, because we think that the non-adiabatic
expansions are not reliable. Therefore, we prefer to extend the present
model later by including the non-adiabatic effects already
in the OBE-potentials.
\item[(iii)]
We extend the OBE-model of \cite{NRS78} further through the inclusion
of phenomenological nucleon-nucleon-meson-meson vertices, henceforth
referred to as 'pair interactions' or 'pair terms'. The vertices are
listed in Table~\ref{table2}.
\end{enumerate}
 
\begin{table}[hbt]
\begin{center}
\vspace{0.5cm}
\begin{tabular}{ccccl} \hline\hline & & & & \\
 $J^{PC}= 0^{++}$ & : &${\cal H}_{S}$ & = &
 $\left(\bar{\psi}'\psi'\right)\left\{
 g_{(\pi\pi)_{0}}\left({\underline \pi}\cdot{\underline \pi}\right)
 + g_{\sigma\sigma} \sigma^{2}\right\}/m_{\pi}$ \\
  & & & & \\
 $J^{PC}= 1^{--}$ & : & ${\cal H}_{V}$ & = &
 $ \left[ g_{(\pi\pi)_{1}}
 \bar{\psi}'\gamma_{\mu}{\underline \tau}\psi'
 - \frac{f_{(\pi\pi)_{1}}}{2M}
 \bar{\psi}'\sigma_{\mu\nu}{\underline \tau}\psi' \partial^{\nu}\right]
 \left({\underline \pi}\times\partial^{\mu}{\underline \pi}\right)
 /m_{\pi}^{2}$ \\
  & & & & \\
  $J^{PC}= 1^{++}$ & : & ${\cal H}_{A}$ & = &
 $g_{(\pi\rho)_{1}} \left(
 \bar{\psi}'\gamma_{\mu}\gamma_{5}{\underline \tau}\psi'\right)\left(
 {\underline \pi}\times{\underline \rho^{\mu}}\right)/m_{\pi}$ \\
  & & & & \\
    & & $ {\cal H}_{P}$ & = &
 $ g_{(\pi\sigma)} \left(
 \bar{\psi}'\gamma_{\mu}\gamma_{5}{\underline \tau}\psi'\right)\left(
 \sigma \partial^{\mu}\underline{\pi} - \underline{\pi}
 \partial^{\mu}\sigma \right)/m_{\pi}^{2}$ \\
  & & & & \\
 $J^{PC}= 1^{+-}$ & : & ${\cal H}_{H}$ & = &
 $ i g_{(\pi\rho)_{0}} \left(
 \bar{\psi}'\sigma_{\mu\nu}\gamma_{5}\psi'\right)\partial^{\nu}\left(
 {\underline \pi}\cdot{\underline \rho^{\mu}}\right)/m_{\pi}^{2}$ \\
  & & & & \\
     & & ${\cal H}_{B}$ & = &
 $ i g_{(\pi\omega)} \left(
 \bar{\psi}'\sigma_{\mu\nu}\gamma_{5}{\underline \tau}
 \psi'\right)\partial^{\nu}\left(
 {\underline \pi}\cdot\omega^{\mu}\right)/m_{\pi}^{2}$ \\
  & & & & \\
 \hline\hline
\end{tabular}
\caption{Phenomenological Meson-Pair Interactions}
\label{table2} \end{center} \end{table}
The motivation for including these 'pair-vertices' is that similar
interactions appear in chiral-lagrangians. They can be viewed upon as
the result of the out integration of the heavy-meson and
resonance degrees of freedom. Moreover, they also represent two-meson
exchange potentials. We are less radical than Weinberg, see
 {\it e.g.} \cite{Weinberg91}, in that we do not integrate out
the degrees of freedom of the mesons with masses below 1 GeV.
The techniques to derive the explicit expressions for
the potentials corresponding to the meson-pair exchange potentials
with soft {\it i.e.} gaussian form factors, is in essence described in
\cite{Rijken91}. The new type of graphs that have to be evaluated are
those with one pair-vertex and with two pair-vertices.
 
Fitting this new model to the NN-data, using
the 1993 Nijmegen single energy $pp + np$ phase shift analysis
\cite{SKRS93}, leads to an excellent result. We reached for the
energies in the range $25 \leq T_{lab} \leq 320$ MeV, which
comprises  3709 data, a $\chi^{2}_{p.d.p.}= 1.16$ \cite{Comment1}.
The (rationalized) coupling constants and form factor masses are
given in Table~\ref{table3}. Here, the $f_{\eta}$ was not fitted but
derived from $f_{\pi}$ using $\alpha_{pv}=0.361$.
We used for the $\sigma$ a mass of $m_{\sigma}= 500.4$ MeV, {\it i.e.}
the lowest mass of the two-pole approximation used in \cite{NRS78}.
The use of different form factors for $I_{t}=1$ and $I_{t}=0$ for vector
and scalar exchange did not have much influence on the fit.
\begin{table}[hbt]
\begin{center}
\vspace{0.5cm}
\begin{tabular}{cc|cc|cc|cc} \hline\hline & & & & & & & \\
 \multicolumn{2}{c|}{ps-pv}   & \multicolumn{2}{c|}{vector} &
   \multicolumn{2}{c|}{scalar} & \multicolumn{2}{c}{pairs}   \\
     &    &    &     &     &     &    &      \\
\hline
     &    &    &     &     &     &    &      \\
 $f_{\pi}$ & 0.268 & $g_{\rho}$  &  0.730 & $g_{\delta}$ &
  1.299     & $g_{(\pi\pi)_{0}}$ &-0.160  \\
     &    &    &     &     &     &    &      \\
 $f_{\eta}$& 0.069    & $f_{\rho}$  &  3.299 & $g_{\epsilon}$ &
  3.573    & $g_{(\pi\pi)_{1}}$ &-0.001 \\
     &    &    &     &     &     &    &      \\
 $f_{\eta'}$ & 0.271   & $g_{\omega}$  &  3.009 & $g_{A_{2}}$ &
  0.123    & $f_{(\pi\pi)_{1}}$ & -0.260 \\
     &      &    &     &     &   &    &      \\
           &         & $f_{\omega}$  &  0.567 & $g_{P}$ &
  2.346    & $g_{(\pi\rho)_{0}}$ &  0.073 \\
     &    &    &     &     &     &    &      \\
 $\Lambda_{PV}$  & 844.8 & $\Lambda_{V,1}$ & 777.6 & $\Lambda_{S,1}$ &
  767.1    & $g_{(\pi\rho)_{1}}$ & 0.506 \\
     &    &    &     &     &     &    &      \\
         &       &$\Lambda_{V,0}$& 744.9 &$\Lambda_{S,0}$ & 835.6 &
   $g_{(\pi\omega) }$ & -0.001 \\
     &    &    &     &     &     &    &      \\
     &    &    &     & $m_{P} $ & 309.1 &
   $g_{(\pi\sigma)}$ &-0.170 \\
     &    &    &     &     &     &    &      \\
         &       &      &      &  &   &
   $g_{(\sigma\sigma)}$ &-0.302 \\
     &    &    &     &     &     &    &      \\
\hline\hline
\end{tabular}
 \caption{Form factor masses, meson and meson-pair couplings.}
\label{table3} \end{center} \end{table}
 
The nuclear-bar phase shifts of the new $N\!N$-model are given in
Table~\ref{table4}. In this table, the $I=1$-phases are $pp$-phases
and the $I=0$-phases are $np$-phases.
The $\chi^{2}$ of the model w.r.t. the
PWA-phases is denoted by $\Delta \chi^{2}$.
 
\begin{table}[hhhhhhhh]
\begin{center}
\begin{tabular}{rrrrrrr} \hline\hline  & & & & & & \\
 $T_{lab}$ & 25 & 50 &100 & 150 & 215 & 320 \\ \hline
     &    &     &     &     &    &      \\
 $\sharp$ data  &352 &572  &399  &676  & 756& 954  \\
     &    &     &     &     &    &      \\
$\Delta \chi^{2}$& 67 & 68 & 30  & 74  & 153 & 335 \\
     &    &     &     &     &    &      \\ \hline
     &    &     &     &     &    &      \\
 $^{1}S_{0}$ & 49.05 & 38.87  & 24.41 & 13.88 & 3.23 & -9.91 \\
 $^{3}S_{1}$ & 79.99 & 62.21  & 42.91 & 30.77 & 19.44 & 6.36 \\
 $\epsilon_{1}$ & 1.96 & 2.38  & 2.82 & 3.21 & 3.70 & 4.43 \\
 $^{3}P_{0}$ & 8.84 & 11.96  & 10.01 & 5.20 &-1.46 &-11.18 \\
 $^{3}P_{1}$ &-4.85 & -8.24  &-13.22 &-17.32 & -21.94 & -28.19 \\
 $^{1}P_{1}$ &-6.19 & -9.39  &-13.87 &-17.81 & -22.57 & -29.40 \\
 $^{3}P_{2}$ & 2.50 &  5.79  & 10.89 & 13.99 &  16.18 &  17.41 \\
 $\epsilon_{2}$ &-0.80 &-1.70  &-2.71 &-3.01 &-2.85 &-2.10 \\
 $^{3}D_{1}$ &-2.78 &-6.39  &-12.27 &-16.71 & -21.17 & -26.56 \\
 $^{3}D_{2}$ & 3.67 & 8.91  & 17.37 & 22.48 &  25.44 &  25.54 \\
 $^{1}D_{2}$ & 0.69 & 1.69  & 3.83  & 5.86  &  7.96  &  9.75  \\
 $^{3}D_{3}$ & 0.05 & 0.31  & 1.34  & 2.53  &  3.77  &  4.70  \\
 $\epsilon_{3}$ & 0.54 & 1.57 & 3.40 & 4.74 & 5.92 & 7.03 \\
 $^{3}F_{2}$ & 0.10 & 0.34  & 0.79  & 1.07  &  1.07  &  0.16  \\
 $^{3}F_{3}$ &-0.22 &-0.66  &-1.47  &-2.11  & -2.82  & -3.97  \\
 $^{1}F_{3}$ &-0.41 &-1.08  &-2.09  &-2.74  & -3.39  & -4.51  \\
 $^{3}F_{4}$ & 0.02 & 0.11  & 0.47  & 0.94  &  1.60  &  2.52  \\
 $\epsilon_{4}$ &-0.05 &-0.19  &-0.52 &-0.82 &-1.13 &-1.49 \\
 $^{3}G_{3}$ &-0.05 &-0.25  &-0.89  &-1.66  & -2.66  & -4.12  \\
 $^{3}G_{4}$ & 0.17 & 0.70  & 2.09  & 3.49  &  5.15  &  7.33  \\
 $^{1}G_{4}$ & 0.04 & 0.15  & 0.41  & 0.67  &  1.03  &  1.63  \\
 $^{3}G_{5}$ &-0.01 &-0.05  &-0.15  &-0.23  & -0.26  & -0.17  \\
 $\epsilon_{5}$ & 0.04 & 0.20  & 0.69 & 1.20 & 1.80 & 2.58 \\
 $^{3}H_{4}$ & 0.00 & 0.03  & 0.11  & 0.21  &  0.36  &  0.56  \\
 $^{3}H_{5}$ &-0.01 &-0.08  &-0.29  &-0.52  & -0.77  & -1.10  \\
 $^{1}H_{5}$ &-0.03 &-0.16  &-0.51  &-0.83  & -1.15  & -1.50  \\
 $^{3}H_{6}$ & 0.00 & 0.01  & 0.04  & 0.10  &  0.21  &  0.44  \\
 $\epsilon_{6}$ &-0.00 &-0.03  &-0.11 &-0.22 &-0.35 &-0.54 \\
     &    &     &     &     &    &      \\
\hline\hline
\end{tabular}
\caption{ESC nuclear-bar $pp$ and $np$ phase shifts in degrees.}
\label{table4} \end{center} \end{table}
The numerical results were obtained by using a coordinate space version
of the model.
The OPEP treatment was adapted to the PWA by
multiplying in momentum space the OPEP of \cite{NRS78} by
$\sqrt{M/E(p)}$-factors for the initial and final state.
 
From Table~\ref{table4} one notices the great improvement of the new
model over the OBE-model \cite{NRS78}. In particular this is obvious
for the $^{1}P_{1}$-, the $^{3}D_{2}$-, and the $^{3}D_{3}$-waves.
The $^{3}F_{2}$-wave however, is bending towards zero too quickly as
a function of energy.
Here the $2\pi$-potential gives a repulsive tensor force.
At this point the inclusion of $\pi\otimes\rho$-, $\pi\otimes\omega$-
potentials in the future may be of help in particular.
The values reported in Table~\ref{table3} are very reasonable. The
pion coupling was searched and $f_{N\!N\pi}^{2}=0.072$, which is on the
lower side of the determinations listed in Table~\ref{table1}. We have
$g_{\rho}^{2}=0.53$ and $(f/g)_{\rho}=4.52$, in reasonable agreement
with VDM \cite{Sakurai68}. The agreement improves if we also
take into account the contribution of the $\pi\pi$-pair terms (see
remark below). The $\omega$-, $\epsilon$-, and pomeron-
couplings are rather similar to those of \cite{NRS78}.
 
\noindent Also the meson-pair couplings are accessible to a physical
interpretation. The couplings $g_{(\pi\pi)_{0}}$ and
$f_{(\pi\pi)_{1}}$ are not very small. This, notwithstanding the fact
that the $I_{t}=0$-channel is dominated by $\epsilon$- and
Pomeron-exchange, which tend to cancel each other largely.
Similarly, because of the dominance of $\rho$-exchange in the
$I_{t}=1$-channel one would tend to expect small values for
$g_{(\pi\pi)_{1}}$ and $f_{(\pi\pi)_{1}}$.
However, the pion-pair contribution represents, among other things,
the correction to the
two-pole approximation used for the description of the broad $\epsilon$
and $\rho$ meson, which is not negligible.
 
\noindent Also, with these $\pi N$-interactions
all s- and p-wave pion-nucleon scattering lengths
are accounted for very well (see also \cite{Padua89}).
In particular, interpreting the $(\pi\pi)_{0}$-pair contribution as
representing in fact the effect of the low mass tail of the broad
$\epsilon$-meson, one finds a contribution $\Delta a_{33} \approx 0.10$,
which is needed together with the nucleon-pole contribution
in order to give the experimental value.
 
\noindent For the
$g_{(\pi\rho)_{1}}$- and $g_{(\pi\sigma)}$-coupling
$A_{1}$-dominance would predict
\begin{eqnarray}
 |g_{(\pi\rho)_{1}}| &=& \left(\frac{m_{\pi}}{m_{A_{1}}}\right)^{2}
  g_{A_{1}N\!N}(0) g_{A_{1}\rho\pi}(0) \approx 0.14 \nonumber \\
  & & \nonumber \\
 |g_{(\pi\sigma)}| &=& \left(\frac{m_{\pi}}{m_{A_{1}}}\right)^{2}
  g_{A_{1}N\!N}(0) g_{A_{1}\sigma\pi}(0) \approx 0.10 \nonumber
\label{eq1} \end{eqnarray}
In obtaining these estimates, we have used the predictions of
the chiral-lagrangians in \cite{Weinberg68} and \cite{Kleinert72}
for $g_{A_{1}\pi\rho}(m_{A_{1}}^{2})$ and
$g_{A_{1}\pi\sigma}(m_{A_{1}}^{2})$. Extrapolation to zero momentum
we have done by using a factor $\exp (- m_{A_{1}}^{2}/{\cal M}^{2})$
, where ${\cal M} = 1$ GeV. Additional input in this estimate is
that $ g_{A_{1}N\!N} \approx (m_{\pi}/m_{A_{1}}) f_{\pi N\!N} = 2.45 $
\cite{Schwinger67} (see also \cite{Kane77}).
Similarly, we find from the chiral-lagrangians the prediction,
using $\sigma$-dominance, that roughly $g_{\sigma\sigma} \approx -0.50$,
which is not far from $-0.30$ found in the fit.
\noindent Likewise, assuming that $g_{(\pi\rho)_{0}}$ and
 $g_{(\pi\omega)}$ are dominated by respectively the H- and
 $B_{1}$-meson, we could estimate from the fitted values the couplings
 $g_{H NN}$ and $g_{B_{1}N\!N} $.
 Of course, heavy boson dominance is not
 valid for all these pair couplings. If we would include also
 the $\pi\otimes\rho$-, $\pi\otimes\omega$- etc.
 potentials, then the residual interactions are more likely to be
 boson dominated. Therefore, the present results are preliminary.
\section{Conclusions and Outlook}
The multi-energy Nijmegen PWA poses a nice new challenge to the
theory of the low momentum transfer baryon-baryon interactions.
The success of our ESC-model indicates
that the better quality of the multi-energy Nijmegen PWA with
respect to other phase shift analyses, indeed opens the door to a
more thorough understanding of the low energy NN-data.
To make progress in the problems concerning Few Body Physics,
it is imperative that baryon-baryon interactions are used which are
based on a very realistic description of nucleon-nucleon scattering.
Conclusions
about such parameters as the pion-nucleon coupling constant,
 the relativistic effects, the off-mass-shell
effects etc. are otherwise liable to be fallacious.
 
The chiral-quark-model picture \cite{Manohar84}
makes it highly implausible that there
will be large nucleon-antinucleon-pair effects in the low energy
region (see also \cite{Swart78}).
Incidentally, a model with large nucleon-antinucleon
pair contributions should also include in the intermediate states
pion-nucleon resonances up to 3 GeV, nucleon-hyperon-kaon intermediate
states etc. Also,
the presence of these pairs in nuclear Compton scattering is improbable.
In fact, it is likely that the negative energy contributions of the
constituents cancel out in the Thomson limit \cite{Brodsky69}.
 
Multi-soft-pion and multi-meson effects on the other hand are expected,
both in chiral-lagrangian models and QCD \cite{Witten79}.
However, for reactions dominated by momentum transfers below 1 GeV,
interactions based on gluon-exchange are presumably
suppressed \cite{Manohar84}. Therefore, models based on strong
gluon-quark exchanges do not seem very realistic.
 
The proper theoretical framework for the phenomenological nucleon-nucleon
meson-pair vertices seems the non-linear chiral $SU(2)\times SU(2)$-
symmetry ( for reference see {\it e.g.}  \cite{Weinberg68}.
Then, the extension from nucleon-nucleon to
baryon-baryon can be tried by employing $SU(3)\times SU(3)$-symmetry
(see {\it e.g.} \cite{SUV69}).
This would introduce only a very restricted set of extra free
parameters in for example hyperon-nucleon models.
 
The extension to higher nucleon-nucleon energies of the
ESC-model requires the explicit treatment of the
$\Delta_{33}$-resonance degrees of freedom. This can be done immediately
and will result in different meson-pair contributions. For low energy
scattering this is unnecessary. This follows on the one hand from our
successful fit to the low energy data by {\it e.g.} the new $N\!N$-model
described above, and on the other hand this is explained to be
possible to a certain degree of accuracy by duality (see the remarks
in \cite{Padua89}).

\section{Acknowledgements}
The generous help of C. Terheggen in the implementation of the
authors programs on the HP9000-735 computer is gratefully acknowledged.
It is also a pleasure to thank the other (former) members of the Nijmegen
group, R. Klomp, J.-L. de Kok, M. Rentmeester, V. Stoks, and
R. Timmermans for many discussions regarding the various aspects of the
multi-energy phase shift analysis. Last but not least, the never fading
interest of J.J. de Swart in baryon-baryon interactions
should be mentioned.

\vspace*{\fill}
 

\begin{thebibliography}{99}
\bibitem{Comment1}
 In the presentation of this work at the conference only the results
 for the MAW-X phase shift analysis \cite{AMcGW69} were available.
 Here, we present exclusively the results of the ESC-model for the PWA
 of reference \cite{SKRS93}.
\end{thebibliography}
\end{document}